\documentclass[10pt,journal]{IEEEtran}
\usepackage{amsmath,amssymb,amsfonts}
\usepackage{graphicx}
\usepackage{cite}
\newcommand{\R}{\mathbb{R}}
\newcommand{\C}{\mathbb{C}}
\newcommand{\Hthree}{H_3}
\newcommand{\ustft}{U-STFT}
\newtheorem{theorem}{Theorem}
\newtheorem{proposition}{Proposition}
\newtheorem{corollary}{Corollary}

\begin{document}

\title{Universal Short-Time Fourier Transform: Data-Adaptive
Window-Length Selection for Time-Frequency Analysis}

\author{
\IEEEauthorblockN{
Manan Mittal\IEEEauthorrefmark{1},
Kanad Sarkar\IEEEauthorrefmark{2},
Ryan M. Corey\IEEEauthorrefmark{3},
John R. Buck\IEEEauthorrefmark{4}, and
Andrew C. Singer\IEEEauthorrefmark{1}
\thanks{The Office of Naval Research supported this work in part
under Award N00014-23-1-2133.}
}
\vspace{0.2cm} \\
\IEEEauthorblockA{
\IEEEauthorrefmark{1}Stony Brook University,
\IEEEauthorrefmark{2}University of Illinois Urbana Champaign,
\IEEEauthorrefmark{3}University of Illinois Chicago,
\IEEEauthorrefmark{4}University of Massachusetts Dartmouth
}
}
\maketitle

\begin{abstract}
The short-time Fourier transform (STFT) characterizes the local frequency content of time-varying signals by evaluating the discrete Fourier transform over a sliding analysis window. The STFT is governed by a fundamental resolution trade-off: a short window localizes rapid transients in time at the cost of frequency resolution, whereas a long window resolves closely spaced harmonics at the expense of temporal resolution. The Universal Short-Time Fourier Transform (\ustft) resolves this trade-off sequentially by modeling the analysis window duration as a data-adaptive state variable that indexes the number of hops included in the current observation interval. At each frame, a bank of candidate window durations is scored using normalized order-three R\'enyi entropy to quantify joint time-frequency concentration. A window length trellis aggregates the resulting $\mathcal{O}(3^M)$ candidate window schedules with linear complexity per frame, guaranteeing a cumulative mix-loss regret bound of $\mathcal{O}((k+1)\log M)$ relative to any hindsight window schedule with $k$ resolution changes over $M$ frames. By linearity of the Fourier transform, posterior expectation across the candidate bank yields a single spectrum equivalent to an STFT evaluated with a data-adaptive effective window, while an envelope-normalized synthesis stream guarantees exact overlap-add reconstruction. Evaluations on synthetic and marine bioacoustic signals demonstrate adaptive resolution tracking, enhanced spectral concentration, and exact time-domain invertibility.
\end{abstract}


\section{Introduction}

The short-time Fourier transform (STFT) maps a one-dimensional discrete-time signal into a two-dimensional representation of time and frequency \cite{gabor1946theory,allen1977short,portnoff1980time,cohen1995time,grochenig2001foundations}. At each analysis frame, the transform multiplies the signal by a finite-duration window in time and evaluates the discrete Fourier transform (DFT) of the windowed segment \cite{allen1977short,mallat2008wavelet}. Because multiplication in the time domain corresponds to periodic convolution in the frequency domain, the STFT spectrum at each frame is the local signal spectrum convolved with the Fourier transform of the analysis window.

\begin{figure}
\centering
\includegraphics[width=0.85\linewidth]
{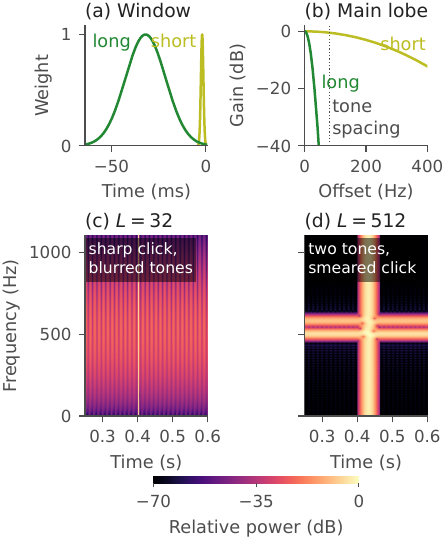}
\caption{The STFT time-frequency resolution trade-off. A short analysis window provides fine temporal localization but coarse frequency resolution due to its wide spectral main lobe. A long analysis window narrows the spectral main lobe for fine frequency resolution at the expense of temporal smearing.}
\label{fig:window-tradeoff}
\end{figure}

The duality between time and frequency has a fundamental trade-off between temporal and spectral resolution governed by the Gabor--Heisenberg uncertainty principle \cite{grochenig2001foundations,cohen1995time}. A short analysis window restricts temporal integration to a brief interval, isolating impulsive events and rapid frequency modulations. However, a short window possesses a wide spectral main lobe, smearing energy across adjacent frequency bins and obscuring closely spaced harmonic components. Conversely, a long analysis window spans multiple signal periods, compressing the spectral main lobe to provide fine frequency resolution, but averages time-domain variations over the duration of the window \cite{grochenig2001foundations,cohen1995time}. Physical signals routinely vary on multiple timescales simultaneously. Because the classical STFT applies a single fixed window duration across the entire record, a window chosen to resolve one signal structure necessarily degrades the resolution of the other (Fig.~\ref{fig:window-tradeoff}).

Existing approaches to mitigating this resolution trade-off follow three main directions. Nonlinear post-processing methods such as reassignment \cite{auger1995improving,chassande2003reassignment} and synchrosqueezing \cite{daubechies2011synchrosqueezed} sharpen spectrograms by reallocating spectral energy toward estimated instantaneous frequency trajectories. While effective for isolated ridges, nonlinear energy reallocation alters the linear projection structure of the STFT and complicates signal synthesis when components intersect. Adaptive window-selection procedures instead optimize local concentration metrics across candidate window lengths \cite{jones1990adaptive,jones1992resolution,baraniuk1993signal,jones1994simple,jones1995adaptive,czerwinski1997adaptive,katkovnik1998instantaneous,kwok2000improved,zhong2010adaptive,pei2012adaptive,mann1995chirplet} or tune window parameters via gradient descent \cite{zhao2021optimizing,leiber2023adaptive,leiber2025optimal}. However, selecting window durations independently at each frame via greedy optimization is susceptible to noise, as minor signal perturbations can induce spurious frame-to-frame fluctuations in window duration during stationary intervals. Nonstationary Gabor frames establish exact reconstruction over nonuniform time-frequency lattices \cite{daubechies2002adaptive,jaillet2007jigsaw,rudoy2010superposition,dorfler2011quilted,balazs2011theory,balazs2013synthesis,holighaus2013cqt}, but typically require offline structural specification or dual-frame inversions across irregular frequency grids.

Online sequential prediction and Bayesian changepoint detection offer a principled framework for tracking time-varying signal parameters without greedy instability \cite{adams2007changepoint,fearnhead2006exact,fearnhead2007online,merhav1998universal,cesa2006prediction,vovk1998game,littlestone1994weighted,freund1997decision,herbster1998tracking,koolen2008combining}. However, conventional run-length algorithms restrict state transitions at each step to a binary choice: either increment the run length by one step to accumulate data since the last changepoint, or reset the run length to zero. While suited to piecewise-stationary parameter estimation, a binary grow-or-reset model cannot represent a sliding analysis window of constant finite duration. Maintaining a constant time-frequency trade-off across a stationary interval requires a third operation that holds the window duration fixed while advancing its temporal support.

This paper introduces the Universal Short-Time Fourier Transform (\ustft), which formulates adaptive time-frequency resolution as a sequential decision problem over a finite bank of candidate window durations. At each frame $m$, the algorithm indexes candidate windows by a discrete memory state $s \in \{0, 1, \dots, s_{\max}\}$ representing the number of hops of length $H$ retained alongside the current hop, yielding physical window durations $L_s = (s+1)H$. Rather than committing greedily to a single window length, the \ustft{} maintains a posterior probability distribution over all candidate states. Each candidate window is scored using the normalized sum of temporal and spectral order-three R\'enyi entropies \cite{renyi1961measures,williams1995entropy,flandrin1994renyi,baraniuk2001measuring,stankovic2001measure,liuni2011entropy,liuni2013automatic,meignen2020renyi}, a dual-domain criterion that penalizes both temporal smearing when $L_s$ is overly long across a transient and spectral main-lobe broadening when $L_s$ is overly short during a time-invariant signal component. Theorem~\ref{thm:entropy-floor} proves a strictly positive lower bound on this joint loss via the discrete uncertainty principle \cite{donoho1989uncertainty,lieb1990integral}. Between consecutive frames, the window state may \emph{maintain} its current duration ($s \to s$), \emph{grow} by one hop ($s \to s+1$) to narrow the spectral main lobe, or \emph{reset} to the single-hop baseline ($s \to 0$) to isolate an impulse. Using a symmetric ternary Dirichlet-multinomial prior \cite{krichevsky1981performance}, Theorem~\ref{thm:regret-bound} proves that the resulting universal mixture achieves a cumulative mix-loss regret of $\mathcal{O}((k+1)\log M)$ relative to any hindsight window schedule with $k$ resolution changes over $M$ frames.

Because distinct window schedules that arrive at the same state $s$ at frame $m$ apply the exact same causal window to the same signal samples, their probability masses merge recursively on a Markovian trellis \cite{rabiner1989tutorial}. Theorem~\ref{thm:state-exactness} proves that this recursion computes the Bayesian mixture over all $\mathcal{O}(3^M)$ admissible schedules in $\mathcal{O}(s_{\max})$ operations per frame. Taking the posterior expectation over the candidate spectra then yields a single complex time-frequency surface. By linearity of the DFT, Proposition~\ref{prop:effective-window} shows that this mixture spectrum equals a standard STFT evaluated with a data-adaptive effective window, while Theorem~\ref{thm:posterior-mix-dominance} proves that posterior filtering strictly dominates predictive weighting by eliminating transient lag. Finally, Theorem~\ref{thm:reconstruction} establishes exact time-domain signal reconstruction via an envelope-normalized overlap-add synthesis stream \cite{griffin1984signal}.

\section{Causal Multi-Resolution Candidate Bank}
\label{sec:model}

\subsection{Causal Frame Geometry and Resolution States}

Consider a real-valued discrete-time signal $x[n] \in \R$ sampled at rate $f_s$. The analysis advances in increments of $H \in \mathbb{N}$ samples per hop, so that frame $m \in \{1, \dots, M\}$ terminates at sample index
\begin{equation}
n_m=mH.
\label{eq:frame-boundary}
\end{equation}
For a uniform DFT size $N \ge H$, the causal observation buffer at frame $m$ collects the $N$ samples immediately preceding $n_m$:
\begin{equation}
\begin{split}
\mathbf x_m=[&x[n_m-N],x[n_m-N+1],\\
&\dots,x[n_m-1]]^T\in\R^N,
\end{split}
\label{eq:lookback}
\end{equation}
with zero-padding applied for indices $n < 0$.

To span a range of time-frequency trade-offs within this buffer, we define a bank of discrete memory states $s \in \mathcal{S} = \{0, 1, \dots, s_{\max}\}$, where $s_{\max} = \lfloor N/H \rfloor - 1$. The state $s$ specifies the number of historical hops included alongside the current hop, corresponding to a physical window support of
\begin{equation}
L_s=(s+1)H
\label{eq:support-definition}
\end{equation}
samples. State $s=0$ analyzes only the $H$ samples of the newest hop ($L_0 = H$), providing maximum temporal localization and the widest spectral main lobe ($\Delta \omega \propto 1/H$). Increasing $s$ extends the window further into the past to incorporate $s$ additional hops, progressively narrowing the spectral main lobe ($\Delta \omega \propto 1/L_s$) at the cost of averaging over a longer temporal interval.

Each state $s$ employs a real nonnegative taper $\mathbf{w}_{L_s} \in \R^N$ supported on the rightmost $L_s$ samples of the observation buffer:
\begin{equation}
w_{L_s}[N-L_s+\ell]>0,\qquad 0\le\ell<L_s,
\end{equation}
with $w_{L_s}[\nu] = 0$ for $0 \le \nu < N - L_s$. In our implementation, each active support uses a Gaussian taper truncated at three standard deviations and normalized to unit sum:
\begin{equation}
\sum_{\nu=0}^{N-1}w_{L_s}[\nu]=1.
\label{eq:window-normalization}
\end{equation}
Aligning every candidate taper to the terminal boundary $n_m$ ensures causal operation and guarantees that all candidates evaluate the signal on a common time reference and uniform $N$-bin frequency grid, permitting direct linear combination of their complex DFT coefficients without phase-shifting corrections.

Multiplying the observation buffer elementwise ($\odot$) by the $s$-th candidate taper yields the windowed frame
\begin{equation}
\mathbf z_{m,s}=\mathbf w_{L_s}\odot\mathbf x_m.
\label{eq:windowed-frame}
\end{equation}
Applying the unitary $N$-point DFT $\mathcal{F}_N$, with entries
\begin{equation}
[\mathcal F_N\mathbf z][k]
=\frac{1}{\sqrt N}\sum_{\nu=0}^{N-1}
z[\nu]e^{-i2\pi k\nu/N},
\quad 0\le k<N,
\label{eq:dft-definition}
\end{equation}
produces the $s$-th candidate spectrum at frame $m$:
\begin{equation}
\mathbf X_{m,s}=\mathcal F_N\mathbf z_{m,s}.
\label{eq:candidate-transform}
\end{equation}

\subsection{Admissible Window Schedules and the Hindsight Oracle}

As the signal evolves across frames $m = 1, \dots, M$, selecting a window duration at each frame defines a state trajectory, or window schedule,
\begin{equation}
\tau=(s_1,s_2,\dots,s_M),\qquad
s_m\in\mathcal{S}.
\label{eq:path-definition}
\end{equation}
Because physical observation memory can expand by at most one hop $H$ per frame, a causal schedule $\tau$ transitions between consecutive frames $m \to m+1$ through one of three operations. It may \emph{maintain} the current state ($s_{m+1} = s_m$), sliding the window forward by $H$ samples at constant duration $L_{s_m}$ to preserve the active time-frequency resolution trade-off. It may \emph{grow} by one hop ($s_{m+1} = s_m + 1$), retaining the existing window start time while appending the new hop to lengthen the window to $L_{s_m}+H$ and sharpen frequency resolution. Finally, it may \emph{reset} to the baseline state ($s_{m+1} = 0$), discarding historical hops and collapsing the window to $L_0 = H$ to isolate an abrupt transient.

Let $\mathcal{T}_M$ denote the set of all admissible schedules of length $M$ obeying these transitions, and let $\mathcal{T}_{k,M} \subset \mathcal{T}_M$ denote the subset of schedules containing at most $k$ non-maintain transitions, which partition the record into $k+1$ constant-resolution segments. We evaluate the sequential performance of the \ustft{} against a hindsight oracle that observes the entire signal record $\widetilde{\mathbf{x}}^M = (\mathbf{x}_1, \dots, \mathbf{x}_M)$ and selects the optimal schedule $\tau^* \in \mathcal{T}_{k,M}$.

\section{Quantifying Time-Frequency Resolution via R\'enyi Entropy}
\label{sec:loss}

\subsection{Dual-Domain Energy Distributions}

To determine which candidate window $L_s$ best matches the local signal structure at frame $m$, we measure how tightly each candidate concentrates signal energy in both time and frequency \cite{williams1995entropy,flandrin1994renyi}. For a non-zero windowed frame $\mathbf{z}_{m,s}$ with total energy
\begin{equation}
E_{m,s}=\|\mathbf z_{m,s}\|_2^2,
\label{eq:frame-energy}
\end{equation}
the normalized temporal energy distribution over its active support of length $L_s$ is the probability mass function (PMF)
\begin{equation}
p_{m,s}[\ell]
=\frac{|z_{m,s}[N-L_s+\ell]|^2}{E_{m,s}},
\quad 0\le\ell<L_s.
\label{eq:time-pmf}
\end{equation}
Similarly, let $\mathcal{K} \subseteq \{0, \dots, N-1\}$ denote the set of retained frequency bins (such as the unique non-redundant bins of a real signal) with cardinality $F = |\mathcal{K}| \ge 2$ and positive quadrature weights $\alpha_k > 0$. The normalized spectral energy distribution across $\mathcal{K}$ forms the PMF
\begin{equation}
q_{m,s}[k]
=\frac{\alpha_k|X_{m,s}[k]|^2}
{\sum_{j\in\mathcal K}\alpha_j|X_{m,s}[j]|^2},
\quad k\in\mathcal K.
\label{eq:frequency-pmf}
\end{equation}

\subsection{Normalized Order-Three R\'enyi Entropy and the Uncertainty Floor}

Whereas second-order variance metrics penalize energy based on distance from a single mean---incorrectly penalizing multi-component signals with well-resolved, widely separated harmonics---R\'enyi entropy measures concentration directly from the peakedness of the distribution \cite{renyi1961measures,baraniuk2001measuring,stankovic2001measure,liuni2011entropy,liuni2013automatic,meignen2020renyi}. For a PMF $v = [v_1, \dots, v_K]^T$, the order-$\alpha$ R\'enyi entropy is $H_\alpha(v) = \frac{1}{1-\alpha}\log_2(\sum_i v_i^\alpha)$. Setting $\alpha = 3$ yields the order-three R\'enyi entropy:
\begin{equation}
\Hthree(v)=-\frac12\log_2\!\left(\sum_i v_i^3\right).
\label{eq:renyi3}
\end{equation}
Cubing the probability masses strongly favors sharp spectral peaks and localized impulses while suppressing diffuse background noise.

To evaluate the joint time-frequency trade-off across windows of varying physical support $L_s$, we normalize the temporal and spectral R\'enyi entropies by their respective uniform-distribution maxima ($\log_2 L_s$ and $\log_2 F$):
\begin{equation}
\ell_{m,s}
=\frac{\Hthree(p_{m,s})}{\log_2 L_s}
+\frac{\Hthree(q_{m,s})}{\log_2 F}.
\label{eq:normalized-score}
\end{equation}
Each term in \eqref{eq:normalized-score} lies in $[0,1]$. The physical intuition behind $\ell_{m,s}$ reflects the STFT resolution trade-off directly: if $L_s$ is too short for a stationary harmonic tone, convolution with the wide window main lobe disperses energy across frequency bins, inflating $\Hthree(q_{m,s})/\log_2 F$. Conversely, if $L_s$ is too long across an impulse or rapid chirp, the extended window incorporates background noise or nonstationary frequency sweeps, inflating $\Hthree(p_{m,s})/\log_2 L_s$ or smearing the spectrum. Minimizing $\ell_{m,s}$ selects the window length that best balances temporal and spectral localization.

Because no nonzero discrete signal can be simultaneously concentrated at a single time sample and a single frequency bin \cite{donoho1989uncertainty,lieb1990integral}, the resolution loss $\ell_{m,s}$ is bounded strictly away from zero.

\begin{theorem}[Positive Discrete STFT Entropy Floor]
\label{thm:entropy-floor}
Consider a fixed DFT size $N$, a candidate support bank $\mathcal L=\{L_s:s\in\mathcal S\}$ with $L_s\ge2$, and either the full $N$-bin DFT on complex frames or the unique-bin DFT on real frames, with positive weights $\alpha_k>0$ for all $k\in\mathcal K$ ($F=|\mathcal K|\ge 2$). There exists a constant $c_{N,\mathcal L,\mathcal K,\boldsymbol\alpha}>0$ such that every non-zero windowed frame satisfies
\begin{equation}
\frac{\Hthree(p_{m,s})}{\log_2 L_s}
+\frac{\Hthree(q_{m,s})}{\log_2 F}
\ge c_{N,\mathcal L,\mathcal K,\boldsymbol\alpha}.
\label{eq:positive-floor}
\end{equation}
\end{theorem}

The proof is given in Appendix~\ref{app:stft-proof}. To ensure a consistent dynamic range across frames with varying signal-to-noise ratios, the raw resolution losses at frame $m$ are standardized across the active candidate set $\mathcal{S}_m$:
\begin{equation}
\begin{split}
d_{m,s}
&=\frac{\ell_{m,s}-\min_{j\in\mathcal S_m}\ell_{m,j}}
{\sigma_m+\epsilon},\\
\sigma_m
&=\operatorname{std}_{j\in\mathcal S_m}(\ell_{m,j}),
\end{split}
\label{eq:standardized-loss}
\end{equation}
where $\epsilon = 10^{-6}$ prevents division by zero when candidate scores coincide. Standardized loss assigns $d_{m,s} = 0$ to the most concentrated candidate at frame $m$ and measures suboptimal windows in units of across-candidate standard deviation.

\section{Sequential Window Tracking and Universal Regret}
\label{sec:regret}

\subsection{Exponential Loss-to-Likelihood Mapping}

Rather than selecting $\arg\min_s d_{m,s}$ greedily at each frame, the \ustft{} embeds the candidate bank within the competitive online prediction framework \cite{merhav1998universal,cesa2006prediction,vovk1998game,littlestone1994weighted,freund1997decision}. For any admissible window schedule $\tau = (s_1, \dots, s_M) \in \mathcal{T}_M$, the cumulative standardized resolution loss over $M$ frames is
\begin{equation}
D_M(\tau)
=\sum_{m=1}^{M}d_{m,s_m}.
\label{eq:path-loss}
\end{equation}
Given an inverse temperature parameter $\beta > 0$, each schedule $\tau$ is assigned the exponential performance weight, or competitive likelihood,
\begin{equation}
P(\widetilde{\mathbf x}^{M}\mid\tau)
\triangleq
\exp\!\left\{-\beta D_M(\tau)\right\}.
\label{eq:path-likelihood}
\end{equation}
The hindsight oracle schedule $\tau^*$ maximizes this weight across all schedules with at most $k$ resolution changes:
\begin{equation}
\begin{split}
P^*(\widetilde{\mathbf x}^{M})
=&\max_{\tau\in\mathcal T_{k,M}} 
P(\widetilde{\mathbf x}^{M}\mid\tau) \\
=& \exp\!\left\{-\beta\min_{\tau\in\mathcal T_{k,M}}D_M(\tau)\right\}.
\end{split}
\label{eq:best-schedule}
\end{equation}

\subsection{Universal Mixture Over Window Schedules}

Because the optimal schedule $\tau^*$ is unknown causally, the universal algorithm assigns a prior probability $P(\tau)$ to every admissible trajectory $\tau \in \mathcal{T}_M$ and forms the prior-weighted mixture
\begin{equation}
\begin{split}
\widetilde P_u(\widetilde{\mathbf x}^{M})
=&\sum_{\tau\in\mathcal T_M}
P(\tau)P(\widetilde{\mathbf x}^{M}\mid\tau) \\
=&\sum_{\tau\in\mathcal T_M}
P(\tau)\exp\!\left\{-\beta D_M(\tau)\right\}.
\end{split}
\label{eq:universal-mixture}
\end{equation}
Defining the cumulative mix-loss of the universal mixture as $L_M^{\mathrm{mix}} \triangleq -\beta^{-1}\ln \widetilde{P}_u(\widetilde{\mathbf{x}}^M)$ and bounding the sum in \eqref{eq:universal-mixture} below by its single term at $\tau^* \in \mathcal{T}_{k,M}$ yields the fundamental mixture domination inequality:
\begin{equation}
L_M^{\mathrm{mix}}
\le D_M(\tau^*) + \frac{1}{\beta}\ln\frac{1}{P(\tau^*)}.
\label{eq:mixture-domination}
\end{equation}
Equation \eqref{eq:mixture-domination} shows that the excess cumulative loss, or regret, of the universal mixture relative to any hindsight schedule $\tau^*$ is bounded above by the negative log-prior code length $-\beta^{-1}\ln P(\tau^*)$ required to specify $\tau^*$.

\subsection{Ternary Dirichlet-Multinomial Prior and Regret Bounds}

To assign prior probabilities $P(\tau)$ that penalize erratic window switching while permitting rapid adaptation when signal statistics change, we model the sequence of ternary transition actions $\mathcal{A} = \{\text{reset}, \text{grow}, \text{maintain}\}$ using a symmetric Dirichlet-multinomial mixture with parameter $1/3$ over the probability simplex $\Delta^2$ \cite{krichevsky1981performance}. For a trajectory accumulating $r$ resets, $g$ growths, and $u$ maintains, the marginal Dirichlet integral is
\begin{equation}
\begin{split}
P_{\text{Dir}}(r,g,u)
\triangleq{}&\frac{\Gamma(1)}{\Gamma(1/3)^3}\\
&\times\int_{\Delta^2}
\theta_r^{r-2/3}\theta_g^{g-2/3}\theta_u^{u-2/3}
\,d\boldsymbol\theta,
\end{split}
\label{eq:dirichlet-integral}
\end{equation}
which induces the ternary Krichevsky--Trofimov (KT) sequential transition rules with add-$1/3$ smoothing:
\begin{align}
P(a_t=\text{reset}\mid\mathbf c)
&=\frac{r+1/3}{r+g+u+1},
\label{eq:kt-reset}\\
P(a_t=\text{grow}\mid\mathbf c)
&=\frac{g+1/3}{r+g+u+1},
\label{eq:kt-grow}\\
P(a_t=\text{maintain}\mid\mathbf c)
&=\frac{u+1/3}{r+g+u+1}.
\label{eq:kt-maintain}
\end{align}
Integrating over the simplex assigns an action sequence of length $M-1$ with action counts $\{c_a\}_{a\in\mathcal{A}}$ the closed-form prior mass
\begin{equation}
P(\tau)
=\pi_0(s_1)\frac{1}{\Gamma(M)}\prod_{a\in\mathcal A}
\frac{\Gamma(c_a+1/3)}{\Gamma(1/3)},
\label{eq:dirichlet-path-probability}
\end{equation}
where $\pi_0(s_1)$ is the initial state prior at frame $m=1$.

\begin{proposition}[Schedule Code Length]
\label{prop:path-code}
Let $\pi_{\min}=\min_s\pi_0(s)>0$ denote the minimum initial prior mass. For any admissible window schedule $\tau \in \mathcal{T}_{k,M}$ containing $k$ non-maintain transitions over $M$ frames, the negative log-prior satisfies
\begin{equation}
-\ln P(\tau)
\le\ln\frac{1}{\pi_{\min}}
+2(k+1)\ln\frac{M}{k+1}+\mathcal O(k).
\label{eq:path-code-bound}
\end{equation}
\end{proposition}

\begin{IEEEproof}
By the minimax redundancy bound for the Krichevsky--Trofimov mixture over a ternary action alphabet ($A=3$) with $k$ switching points partitioning $M$ steps into $k+1$ segments \cite{krichevsky1981performance,cesa2006prediction,herbster1998tracking,koolen2008combining}, the negative log-probability is bounded by $\ln(1/\pi_{\min}) + \frac{A+1}{2}(k+1)\ln\frac{M}{k+1} + \mathcal{O}(k)$. Substituting $A=3$ yields the prefactor $2(k+1)$ in \eqref{eq:path-code-bound}.
\end{IEEEproof}

Combining Proposition~\ref{prop:path-code} with the mixture domination bound \eqref{eq:mixture-domination} establishes the universal regret guarantee of the \ustft{}.

\begin{theorem}[U-STFT Mix-Loss Regret]
\label{thm:regret-bound}
For any observation sequence $\widetilde{\mathbf x}^{M}$ and any hindsight window schedule $\tau^* \in \mathcal{T}_{k,M}$ with $k$ non-maintain transitions, the cumulative mix-loss regret of the \ustft{} satisfies
\begin{equation}
\begin{split}
\mathcal R_M(k) \triangleq& L_M^{\mathrm{mix}} - D_M(\tau^*) \\
\le&\beta^{-1}\left(
\ln\frac{1}{\pi_{\min}} + 2(k+1)\ln\frac{M}{k+1}+\mathcal O(k)
\right).
\end{split}
\label{eq:regret-theorem}
\end{equation}
\end{theorem}

\begin{IEEEproof}
Substituting the code-length upper bound \eqref{eq:path-code-bound} for $\tau^*$ directly into the mixture domination inequality \eqref{eq:mixture-domination} establishes \eqref{eq:regret-theorem}.
\end{IEEEproof}

When compared against any classical fixed-window STFT, which corresponds to a static schedule ($k=0$) that maintains a single state $s^*$ across all $M$ frames, Theorem~\ref{thm:regret-bound} specializes to a vanishing per-frame regret.

\begin{corollary}[Asymptotic Optimality Relative to Fixed-Window STFTs]
\label{cor:fixed-window}
For any fixed window state $s^*\in\mathcal S$ initialized with $\pi_0(s^*)>0$, the cumulative excess loss satisfies
\begin{equation}
\begin{split}
L_M^{\mathrm{mix}}-\sum_{m=1}^{M}d_{m,s^*}
\le\beta^{-1}\left(
\ln\frac{1}{\pi_0(s^*)}+2\ln M+\mathcal O(1)
\right),
\end{split}
\label{eq:fixed-window-bound}
\end{equation}
so that the time-averaged excess resolution loss vanishes as $M \to \infty$:
\begin{equation}
\lim_{M\to\infty}\frac{1}{M}
\left(L_M^{\mathrm{mix}}-\sum_{m=1}^{M}d_{m,s^*}\right)=0.
\end{equation}
\end{corollary}

\section{Markovian Trellis, Effective Window, and Exact Synthesis}
\label{sec:algorithm}
\label{sec:synthesis}

\subsection{Exact State Aggregation on the Memory Trellis}

Direct summation over all $\mathcal{O}(3^M)$ schedules in \eqref{eq:universal-mixture} is computationally intractable. However, the causal geometry of the candidate bank induces a critical Markovian reduction: any two schedules that arrive at the same memory state $s$ at frame $m$ apply the exact same physical window $\mathbf{w}_{L_s}$ to the observation buffer $\mathbf{x}_m$, producing identical candidate spectra $\mathbf{X}_{m,s}$ and identical losses $d_{m,s}$.

\begin{figure}
\centering
\includegraphics[width=0.85\linewidth]
{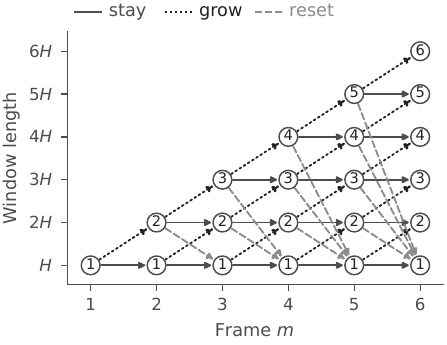}
\caption{Causal window-length trellis. State $s$ corresponds to a window duration $L_s=(s+1)H$. All trajectories arriving at state $s$ at frame $m$ share the same physical window $\mathbf{w}_{L_s}$ and candidate spectrum $\mathbf{X}_{m,s}$, collapsing $\mathcal{O}(3^M)$ schedules into $\mathcal{O}(s_{\max})$ recursive state updates per frame.}
\label{fig:window-trellis}
\end{figure}

Because state $s$ represents $g = s$ cumulative growth steps since the most recent reset, conditioning the ternary KT transitions \eqref{eq:kt-reset}--\eqref{eq:kt-maintain} on the current run depth ($g=s$, $u=r=0$) yields state-dependent Markovian transition probabilities:
\begin{align}
P_T(s\mid s)
&=\frac{1/3}{s+1}
&&\text{(maintain)},
\label{eq:trans-maintain}\\
P_T(s+1\mid s)
&=\frac{s+1/3}{s+1}
&&\text{(grow)},
\label{eq:trans-grow}\\
P_T(0\mid s)
&=\frac{1/3}{s+1}
&&\text{(reset)}.
\label{eq:trans-reset}
\end{align}
At the maximum buffer capacity $s = s_{\max}$, further window growth is inadmissible, and the transition distribution reduces to a binary KT update between maintaining $s_{\max}$ and resetting to $s=0$:
\begin{equation}
\begin{split}
P_T(s_{\max}\mid s_{\max})
&=\frac{s_{\max}+1/2}{s_{\max}+1},\\
P_T(0\mid s_{\max})
&=\frac{1/2}{s_{\max}+1}.
\end{split}
\label{eq:boundary-transition}
\end{equation}
Notice the physical behavior encoded by \eqref{eq:trans-maintain}--\eqref{eq:trans-reset}: at short window lengths (small $s$), the probabilities of maintaining, growing, and resetting are comparable, allowing agile adaptation during transients. As a harmonic component persists and the window grows (large $s$), $P_T(s+1 \mid s) = (s+1/3)/(s+1)$ approaches unity, promoting progressive spectral sharpening until a transient incurs a high temporal entropy penalty and triggers a reset to $s=0$ (Fig.~\ref{fig:window-trellis}). For applications requiring large $s_{\max}$ with minimal computation, the trellis can also be restricted to a sparse dyadic lattice
\begin{equation}
\mathbb L_{\text{dyadic}}
=\{0,1,2,4,8,\dots,s_{\max}\},
\label{eq:dyadic-lattice}
\end{equation}
reducing complexity from $\mathcal{O}(s_{\max})$ to $\mathcal{O}(\log_2 s_{\max})$ candidate DFTs per frame by matching expected transition dwell times across octave steps.

Let $\pi_{m-1}(s)$ denote the predictive prior probability of state $s$ before observing frame $m$, initialized uniformly as $\pi_0(s) = 1/(s_{\max}+1)$. Upon observing $\mathbf{x}_m$ and evaluating the standardized concentration losses $d_{m,s}$, each state receives the local likelihood
\begin{equation}
P_m(\mathbf x_m\mid\widetilde{\mathbf x}^{m-1},s)
=\exp\{-\beta d_{m,s}\}.
\label{eq:predictive-likelihood}
\end{equation}
The marginal frame evidence is the prior-weighted expectation
\begin{equation}
\begin{split}
\widetilde P_u(\mathbf x_m\mid\widetilde{\mathbf x}^{m-1})
={}&\sum_{s=0}^{s_{\max}}\pi_{m-1}(s)\\
&\quad\times
P_m(\mathbf x_m\mid\widetilde{\mathbf x}^{m-1},s),
\end{split}
\label{eq:forward-evidence}
\end{equation}
and Bayes' rule yields the filtered posterior distribution over window durations at frame $m$:
\begin{equation}
\mu_m(s)
=\frac{\pi_{m-1}(s)
P_m(\mathbf x_m\mid\widetilde{\mathbf x}^{m-1},s)}
{\widetilde P_u(\mathbf x_m\mid\widetilde{\mathbf x}^{m-1})}.
\label{eq:forward-posterior}
\end{equation}
Propagating $\mu_m(s)$ through the Markovian transition kernel forms the predictive prior for frame $m+1$:
\begin{equation}
\pi_m(0)
=\sum_{j=0}^{s_{\max}}\mu_m(j)P_T(0\mid j),
\label{eq:prior-reset}
\end{equation}
\begin{equation}
\begin{aligned}
\pi_m(s)
&=\mu_m(s)P_T(s\mid s)
+\mu_m(s-1)P_T(s\mid s-1),\\
&\qquad 0<s\le s_{\max}.
\end{aligned}
\label{eq:prior-step}
\end{equation}

\begin{theorem}[Exact State Aggregation]
\label{thm:state-exactness}
The forward trellis recursion \eqref{eq:predictive-likelihood}--\eqref{eq:prior-step} computes the exact universal mixture probability over all admissible schedules $\mathcal{T}_M$ in $\mathcal{O}(s_{\max})$ operations per frame:
\begin{equation}
\begin{split}
\prod_{m=1}^{M}
\widetilde P_u(\mathbf x_m\mid\widetilde{\mathbf x}^{m-1})
=\sum_{\tau\in\mathcal T_M}
P(\tau)\exp\!\left\{-\beta D_M(\tau)\right\}.
\end{split}
\label{eq:state-exactness}
\end{equation}
\end{theorem}

\begin{IEEEproof}
We proceed by induction on $M$. For $M=1$, $\widetilde{P}_u(\mathbf{x}_1) = \sum_{s_1} \pi_0(s_1) e^{-\beta d_{1,s_1}}$, matching the right-hand side of \eqref{eq:state-exactness}. Suppose the product up to $m-1$ equals the sum over all prefix trajectories $\tau_{1:m-1}$, and that $\pi_{m-1}(s)\prod_{j=1}^{m-1}\widetilde{P}_u(\mathbf{x}_j \mid \widetilde{\mathbf{x}}^{j-1})$ equals the sum of prior-weighted likelihoods of all prefixes extending to state $s$ at frame $m$. Multiplying by $e^{-\beta d_{m,s}}$ in \eqref{eq:forward-evidence}--\eqref{eq:forward-posterior} incorporates the frame-$m$ loss, and applying the transition updates \eqref{eq:prior-reset}--\eqref{eq:prior-step} distributes the mass across all valid extensions $s_{m+1}$ via the Markovian kernel $P_T(s_{m+1} \mid s_m)$ \cite{rabiner1989tutorial}. Summing over $s_M$ at frame $M$ completes the induction.
\end{IEEEproof}

\subsection{Effective Analysis Window, Stationary Phase, and Posterior Dominance}

Rather than selecting a single candidate spectrum, the \ustft{} forms its composite complex spectrum $S_A[m,k]$ by taking the posterior expectation over the candidate bank:
\begin{equation}
S_A[m,k]
=\sum_{s=0}^{s_{\max}}\mu_m(s)X_{m,s}[k].
\label{eq:complex-blend}
\end{equation}
Defining the posterior-weighted effective analysis window
\begin{equation}
\mathbf g_m
=\sum_{s=0}^{s_{\max}}\mu_m(s)\mathbf w_{L_s},
\label{eq:effective-window}
\end{equation}
linearity of the DFT guarantees that spectral mixture averaging is equivalent to time-domain adaptive windowing.

\begin{proposition}[Effective-Window Identity]
\label{prop:effective-window}
The posterior-weighted composite spectrum \eqref{eq:complex-blend} is identical to a standard DFT evaluated with the data-adaptive effective window $\mathbf{g}_m$:
\begin{equation}
S_A[m,\cdot]
=\mathcal F_N(\mathbf x_m\odot\mathbf g_m).
\label{eq:effective-window-identity}
\end{equation}
\end{proposition}

\begin{IEEEproof}
Substituting $\mathbf{X}_{m,s} = \mathcal{F}_N(\mathbf{w}_{L_s} \odot \mathbf{x}_m)$ into \eqref{eq:complex-blend} and invoking the linearity of $\mathcal{F}_N$ yields
\begin{equation}
\begin{split}
S_A[m,\cdot]
&=\sum_{s=0}^{s_{\max}}\mu_m(s)
\mathcal F_N(\mathbf w_{L_s}\odot\mathbf x_m)\\
&=\mathcal F_N\!\left(
\left[\sum_{s=0}^{s_{\max}}\mu_m(s)\mathbf w_{L_s}\right]
\odot\mathbf x_m\right) \\
&=\mathcal F_N(\mathbf g_m\odot\mathbf x_m).
\end{split}
\end{equation}
\end{IEEEproof}

Proposition~\ref{prop:effective-window} reveals why coherent spectral averaging sharpens time-frequency ridges without introducing nonlinear cross-terms. By the principle of stationary phase \cite{bleistein1986asymptotic}, for an oscillatory component $A(t)e^{i\Phi(t)}$, the continuous counterpart of the candidate transform is
\begin{equation}
\begin{split}
X_{m,s}(\omega)\approx
\int{}&A(n_m-N+u)w_{L_s}(u)\\
&\times e^{i\Psi_m(u;\omega)}\,du,
\end{split}
\label{eq:stationary-integral}
\end{equation}
where $\Psi_m(u;\omega) = \Phi(n_m - N + u) - \omega u$. Along an instantaneous frequency ridge where $\partial \Psi_m / \partial u = 0$, the phase is stationary and aligned across all nonnegative candidate windows $\mathbf{w}_{L_s}$, so their complex weighted sum \eqref{eq:complex-blend} interferes constructively. Away from the ridge, rapid phase rotation induces destructive cancellation across differing window supports.

Furthermore, because causal time-frequency analysis forms the spectrum of frame $m$ \emph{after} observing the current hop $\mathbf{x}_m$, the \ustft{} synthesizes $S_A[m,k]$ using the filtered posterior $\mu_m(s)$ rather than the predictive prior $\pi_{m-1}(s)$. Conditioning on $\mathbf{x}_m$ eliminates one-step switching delay at transient boundaries and strictly improves upon the predictive mix-loss.

\begin{theorem}[Posterior Mix-Loss Dominance]
\label{thm:posterior-mix-dominance}
Let $d_m^{\mathrm{mix}} \triangleq -\beta^{-1} \ln \widetilde{P}_u(\mathbf{x}_m \mid \widetilde{\mathbf{x}}^{m-1})$ denote the predictive mix-loss at frame $m$, and let $\bar{d}_m^{\mathrm{post}} \triangleq \sum_{s=0}^{s_{\max}} \mu_m(s) d_{m,s}$ denote the posterior-expected loss. For any observation sequence $\widetilde{\mathbf{x}}^M$ and $\beta > 0$,
\begin{equation}
\label{eq:post-kl-identity}
\bar{d}_m^{\mathrm{post}} = d_m^{\mathrm{mix}} - \frac{1}{\beta} D_{\mathrm{KL}}(\mu_m \,\|\, \pi_{m-1}) \le d_m^{\mathrm{mix}},
\end{equation}
where $D_{\mathrm{KL}}(\mu_m \,\|\, \pi_{m-1}) = \sum_{s=0}^{s_{\max}} \mu_m(s) \ln \frac{\mu_m(s)}{\pi_{m-1}(s)} \ge 0$ is the Kullback--Leibler divergence from the predictive prior to the filtered posterior. Summing over $m=1,\dots,M$ and applying Theorem~\ref{thm:regret-bound} yields
\begin{equation}
\label{eq:post-cumulative-regret}
\begin{split}
\sum_{m=1}^{M} \bar{d}_m^{\mathrm{post}} + \frac{1}{\beta}\sum_{m=1}^{M} D_{\mathrm{KL}}(\mu_m \,\|\, \pi_{m-1}) = L_M^{\mathrm{mix}} \\\le \min_{\tau \in \mathcal{T}_{k,M}} D_M(\tau) + \mathcal{R}_M(k),
\end{split}
\end{equation}
and whenever $d_m(S_A) - \bar{d}_m^{\mathrm{post}} \le \beta^{-1} D_{\mathrm{KL}}(\mu_m \,\|\, \pi_{m-1})$, the realized composite spectrum satisfies $\sum_{m=1}^{M} d_m(S_A) \le L_M^{\mathrm{mix}}$.
\end{theorem}

\begin{IEEEproof}
From the Bayes update \eqref{eq:forward-posterior}, $\pi_{m-1}(s) e^{-\beta d_{m,s}} = \mu_m(s) \widetilde{P}_u(\mathbf{x}_m \mid \widetilde{\mathbf{x}}^{m-1})$. Taking logarithms and rearranging yields $\ln(\mu_m(s)/\pi_{m-1}(s)) = -\beta d_{m,s} + \beta d_m^{\mathrm{mix}}$. Taking the expectation under $\mu_m(s)$ gives
\begin{align}
D_{\mathrm{KL}}(\mu_m \,\|\, \pi_{m-1}) &= \sum_{s=0}^{s_{\max}} \mu_m(s) \ln \frac{\mu_m(s)}{\pi_{m-1}(s)} \nonumber \\
&= -\beta \bar{d}_m^{\mathrm{post}} + \beta d_m^{\mathrm{mix}}.
\end{align}
Dividing by $\beta > 0$ proves \eqref{eq:post-kl-identity}, and $D_{\mathrm{KL}}(\mu_m \,\|\, \pi_{m-1}) \ge 0$ implies $\bar{d}_m^{\mathrm{post}} \le d_m^{\mathrm{mix}}$. Summing over $m=1,\dots,M$ yields \eqref{eq:post-cumulative-regret}.
\end{IEEEproof}

\subsection{Companion Synthesis Stream and Exact Invertibility}

In a conventional fixed-window STFT, unweighted overlap-add synthesis reconstructs $x[n]$ up to a constant scale factor whenever the shifted windows satisfy the constant-overlap-add (COLA) condition \cite{allen1977short,portnoff1980time}. Because the effective window $\mathbf{g}_m$ of the \ustft{} adapts at each frame in response to the signal's time-frequency structure, the sum of overlapping windows varies with time $n$.

To guarantee exact invertibility without iterative synthesis approximations \cite{griffin1984signal}, the \ustft{} maintains an envelope-normalized companion synthesis stream. Let $g_m^{\mathrm{ext}}[n]$ denote the effective window $\mathbf{g}_m$ placed at its global time alignment ending at $n_m = mH$. The cumulative sample envelope is
\begin{equation}
E[n]=\sum_m g_m^{\mathrm{ext}}[n].
\label{eq:envelope}
\end{equation}
Because every candidate window is strictly positive over the most recent hop of length $H$ and $\sum_s \mu_m(s) = 1$, $E[n]$ is strictly positive across the entire analyzed signal record. Pre-normalizing the input signal by this causal sample envelope yields
\begin{equation}
u[n]=\frac{x[n]}{E[n]},
\label{eq:prenormalized-signal}
\end{equation}
evaluated whenever $E[n] > \epsilon_E = 10^{-12}$. Extracting the frame buffer $\mathbf{u}_m \in \R^N$ of $u[n]$ at frame $m$ and applying the effective window $\mathbf{g}_m$ produces the companion synthesis spectrum:
\begin{equation}
S_F[m,\cdot]
=\mathcal F_N(\mathbf u_m\odot\mathbf g_m).
\label{eq:filtering-stream}
\end{equation}

\begin{theorem}[Exact Identity Reconstruction]
\label{thm:reconstruction}
Assuming exact arithmetic and $E[n] > 0$ across the analysis interval, applying the inverse unitary DFT $\mathcal{F}_N^{-1}$ to the unmodified companion coefficients $S_F[m,\cdot]$ followed by unweighted overlap-add synthesis reconstructs $x[n]$ identically.
\end{theorem}

\begin{IEEEproof}
Applying $\mathcal{F}_N^{-1}$ to \eqref{eq:filtering-stream} recovers the windowed pre-normalized frame
\begin{equation}
\mathbf z_m^F
=\mathcal F_N^{-1}S_F[m,\cdot]
=\mathbf u_m\odot\mathbf g_m.
\end{equation}
Summing the aligned frames via unweighted overlap-add yields
\begin{equation}
\begin{split}
\widehat x[n]
&=\sum_m u[n]g_m^{\mathrm{ext}}[n]\\
&=u[n]\sum_m g_m^{\mathrm{ext}}[n]
=u[n]E[n]=x[n].
\end{split}
\end{equation}
\end{IEEEproof}

\section{Experimental Evaluation}

\begin{figure}
\centering
\includegraphics[width=0.95\linewidth]
{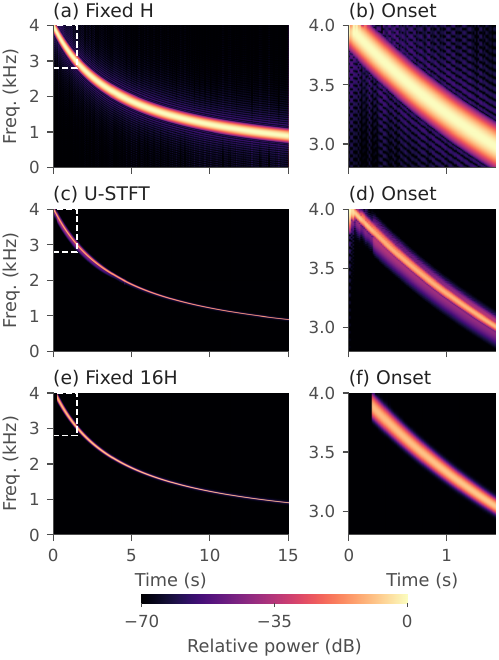}
\caption{Spectrogram comparison for the hyperbolic chirp on a shared decibel scale. The shortest fixed window ($L_0=H$) resolves the rapid onset but broadens the quasi-stationary tail; the longest fixed window ($L_{15}=16H$) sharpens the tail but smears the rapid onset. The \ustft{} adapts its effective window duration continuously to maintain ridge concentration throughout. The right column enlarges the first 1.5~s.}
\label{fig:chirp-comparison}
\end{figure}

\begin{figure*}
\centering
\includegraphics[width=\textwidth]
{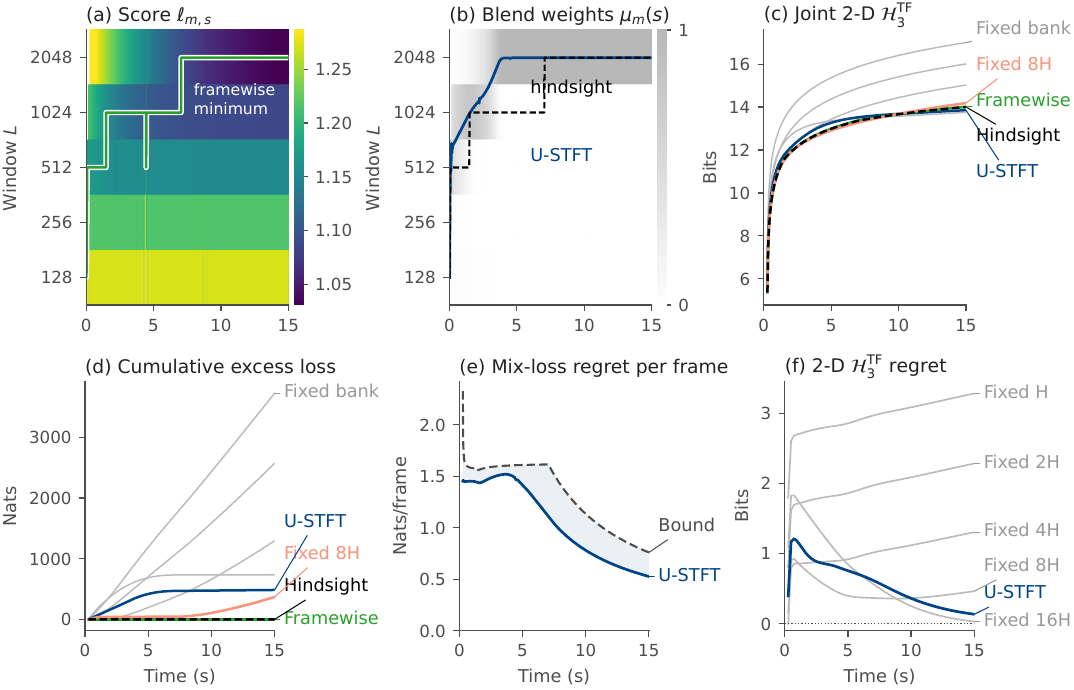}
\caption{Diagnostic trajectories for the hyperbolic chirp: (a) local R\'enyi resolution losses $\ell_{m,s}$ with the framewise minimum, (b) posterior window distribution $\mu_m(s)$ with the \ustft{} and hindsight schedules, (c) global 2-D R\'enyi entropy $\mathcal{H}_3^{\mathrm{TF}}$, (d) cumulative excess loss over the framewise minimum, (e) per-frame mix-loss regret relative to the theoretical bound, and (f) 2-D R\'enyi regret relative to the per-prefix optimal window schedule. Grey curves show the fixed-window bank.}
\label{fig:chirp-diagnostics}
\end{figure*}
\label{sec:experiments}

We evaluate the capacity of the \ustft{} to navigate the time-frequency resolution trade-off across four progressive regimes: a continuous hyperbolic chirp whose instantaneous frequency rate varies smoothly from rapid to quasi-stationary, a noisy composite signal combining frequency sweeps, closely spaced harmonic tones, and abrupt impulses, recorded English speech, and underwater field recordings of marine bioacoustic signals.

Performance is assessed across local resolution, global concentration, and synthesis fidelity. Local resolution is measured by the peak-to-valley decibel separation between closely spaced tones and the equivalent temporal envelope width of impulses. Global spectrogram concentration is quantified via the two-dimensional order-three R\'enyi entropy
\begin{equation}
\mathcal H_3^{\mathrm{TF}}(S)
=-\frac12\log_2\sum_{m,k}
\left(
\frac{|S[m,k]|^2}{\sum_{m',j}|S[m',j]|^2}
\right)^3,
\label{eq:global-renyi}
\end{equation}
where lower values indicate tighter energy concentration across the time-frequency plane. Synthesis fidelity is measured by the normalized reconstruction error $\mathcal{E}_{\mathrm{rec}} = \|x - \widehat{x}\|_2 / \|x\|_2$ computed from the unmodified companion stream $S_F$. Across all trials, we compare the \ustft{} against fixed-window STFT baselines at states $s \in \{0, 7, 15\}$ ($L_s \in \{H, 8H, 16H\}$), greedy framewise selection ($\arg\min_s \ell_{m,s}$), and the hindsight oracle bound.

\subsection{Hyperbolic Chirp: Continuous Resolution Adaptation}

Consider an $8$-kHz sampled hyperbolic chirp whose instantaneous frequency decelerates over $15$~s according to
\begin{equation}
f_c(t)=\frac{4000}{1+(23/101)t}\ \text{Hz},
\qquad 0\le t\le15\ \text{s},
\label{eq:hyperbolic-chirp}
\end{equation}
sweeping from $4.0$~kHz down to $0.906$~kHz. During the initial rapid sweep ($t < 3$~s), a long window spans a wide frequency excursion, smearing energy across time and frequency; tracking the onset requires a short window with fine temporal resolution. Conversely, as the chirp flattens into a slowly varying tone ($t > 8$~s), a short window suffers from a wide spectral main lobe, whereas a long window narrows the ridge.

Figure~\ref{fig:chirp-comparison} demonstrates this resolution trade-off directly. The short fixed window ($L_0 = H$) tracks the steep initial descent cleanly but produces a wide ridge along the tail, whereas the long fixed window ($L_{15} = 16H$) achieves a razor-sharp tail at the expense of severe temporal smearing during the onset. The \ustft{} transitions smoothly from short to long effective windows as the chirp rate decelerates, preserving a concentrated ridge across the entire record.

Figure~\ref{fig:chirp-diagnostics} illustrates the underlying sequential inference. As the sweep rate diminishes, the minimum of the joint entropy loss $\ell_{m,s}$ migrates progressively from small $s$ toward $s_{\max}$. Guided by the grow transition prior, the posterior mass $\mu_m(s)$ tracks this migration smoothly without frame-to-frame oscillations. The empirical per-frame regret stabilizes at $0.526$~nats/frame---well below the theoretical path-code upper bound of $0.760$~nats/frame---and the \ustft{} ends with a lower global R\'enyi entropy $\mathcal{H}_3^{\mathrm{TF}}$ than the best fixed window, greedy framewise selection, and the hindsight schedule.

\subsection{Simultaneous Tone Separation and Transient Localization}

To test how the ternary transition model handles conflicting resolution demands within a single noisy signal, we construct a synthetic waveform embedded in $0$~dB SNR white Gaussian noise containing a rapid linear frequency sweep, two closely spaced stationary sinusoids at $500$~Hz and $532$~Hz, and two broadband impulses at $t = 2$~s and $t = 3$~s.

\begin{figure}
\centering
\includegraphics[width=0.95\linewidth]
{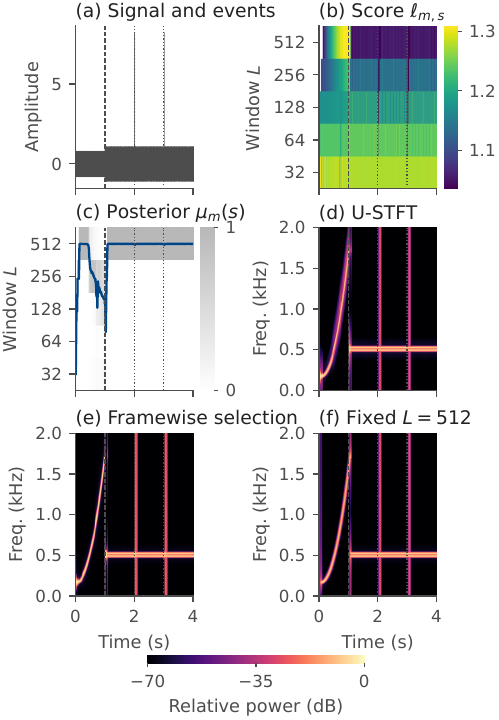}
\caption{Controlled tone and transient experiment at $0$~dB SNR. The \ustft{} grows its window duration to resolve the $500$~Hz and $532$~Hz tones, maintains that duration across the stationary interval, and executes immediate reset transitions ($s \to 0$) at $t=2$~s and $t=3$~s to localize the broadband impulses.}
\label{fig:controlled-evidence}
\end{figure}

Figure~\ref{fig:controlled-evidence} highlights the distinct physical roles of the three trellis transitions. Upon entering the two-tone regime, the trellis executes sequential \emph{grow} transitions, narrowing the spectral main lobe until the $32$-Hz separation between the $500$~Hz and $532$~Hz components is cleanly resolved, and then \emph{maintains} that window length. At the exact onset of each impulse ($t=2$~s and $t=3$~s), the surge in temporal entropy triggers an immediate \emph{reset} transition to $s=0$ ($L_0 = H$), isolating the impulse in time before re-growing the window to resume harmonic tracking. Across six independent noise realizations at $0$~dB SNR, the \ustft{} matches the tonal peak-to-valley separation of the longest fixed window ($L_{15}$) while matching the sharp impulse width of the shortest fixed window ($L_0$). By contrast, greedy framewise selection switches window lengths nearly five times as frequently, introducing severe vertical striping artifacts across the spectrogram.

\subsection{English Speech: Voiced Harmonics and Stop Bursts}

Speech alternates between quasi-stationary voiced segments, whose closely spaced pitch harmonics demand fine spectral resolution, and stop releases and fricative onsets that last only a few milliseconds. We analyze a five-second excerpt of read English speech from speaker p227 of the VCTK corpus \cite{yamagishi2019vctk}, resampled to $16$~kHz, with hop $H = 32$ samples ($2$~ms), so the candidate windows span $L = H$ ($2$~ms) to $L = 16H$ ($32$~ms).

\begin{figure}
\centering
\includegraphics[width=0.95\linewidth]
{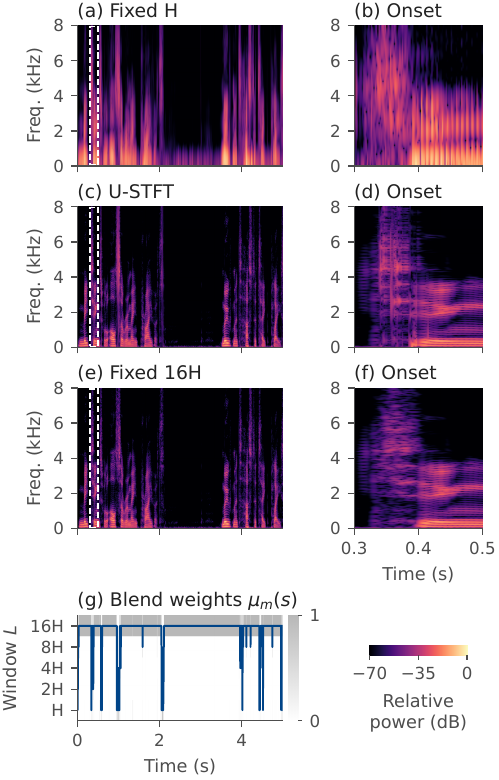}
\caption{Spectrogram comparison for five seconds of English speech (VCTK speaker p227) on a shared decibel scale: (a) the shortest fixed window ($L=H$), (c) the \ustft{}, and (e) the longest fixed window ($L=16H$); (b), (d), and (f) enlarge two stop bursts and the following vowel onset between $0.3$ and $0.5$~s. (g) Posterior window distribution $\mu_m(s)$ with the dominant \ustft{} window. The \ustft{} holds the long window through voiced speech and resets to short windows at broadband onsets.}
\label{fig:speech-comparison}
\end{figure}

Figure~\ref{fig:speech-comparison} exposes the same trade-off within a single utterance. The short fixed window ($L = H$) localizes every burst but merges the voiced harmonics into broad vertical bands, whereas the long fixed window ($L = 16H$) resolves the individual harmonics and formant trajectories at the expense of spreading each burst over $32$~ms. The \ustft{} holds $L = 16H$ on $93\%$ of frames and executes \emph{reset} transitions at broadband onsets such as stop releases, recovering the sharp burst edges of the short window while retaining the harmonic structure of the long window (Fig.~\ref{fig:speech-comparison}(d)). Because voiced segments carry most of the signal energy, the \ustft{} ends with a global R\'enyi entropy $\mathcal{H}_3^{\mathrm{TF}}$ of $12.27$~bits, within $0.11$~bits of the longest fixed window and below greedy framewise selection ($12.51$~bits), which switches window lengths more than four times as frequently.

\subsection{Marine Bioacoustic Field Recordings}

Finally, we evaluate the \ustft{} on five-second underwater acoustic recordings of humpback whale vocalizations (\textit{Megaptera novaeangliae}) \cite{payne1971songs,au2008principles} and snapping shrimp cavitation pulses (\textit{Alpheus heterochaelis}) \cite{versluis2000snapping,ferguson2001situ,noaa2022sanctsound}. These recordings combine prolonged harmonic calls, rapid frequency inflections, and sub-millisecond broadband impulses in ambient ocean noise.

\begin{figure}
\centering
\includegraphics[width=0.95\linewidth]
{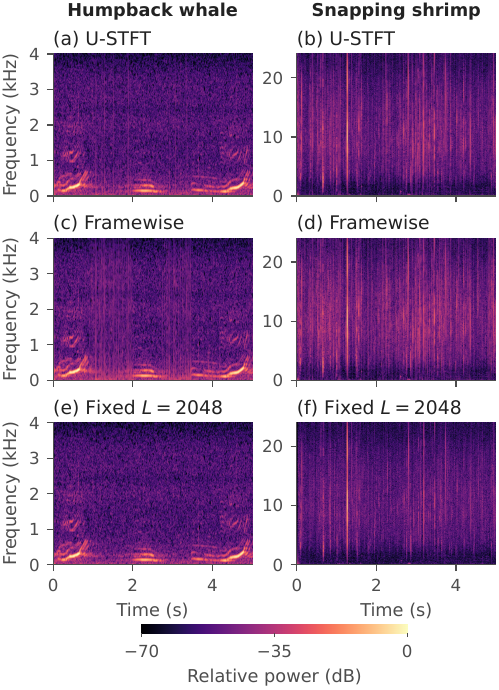}
\caption{Spectrogram comparison on measured marine bioacoustic recordings: humpback whale vocalizations (left) and snapping shrimp cavitation impulses (right) analyzed via the \ustft{}, greedy framewise selection, and fixed-window STFT baselines.}
\label{fig:field-comparison}
\end{figure}

In the humpback whale recording (Fig.~\ref{fig:field-comparison}, left), the \ustft{} expands its effective window during sustained harmonic song units to resolve the overtone series, then contracts the window during rapid frequency modulations to prevent cross-frame smearing. In the snapping shrimp recording (Fig.~\ref{fig:field-comparison}, right), asynchronous cavitation snaps demand fine temporal resolution; the \ustft{} repeatedly resets to $L_0 = H$ at pulse onsets, avoiding the temporal smearing exhibited by long fixed windows and the erratic background artifacts of greedy selection. Across all synthetic, speech, and field trials, companion-stream synthesis reconstructs the time-domain waveforms with normalized error $\mathcal{E}_{\mathrm{rec}} \approx 10^{-16}$, confirming exact invertibility to double-precision machine epsilon.

\section{Conclusion}

The classical STFT imposes a compromise between temporal localization and spectral main-lobe resolution. The Universal Short-Time Fourier Transform (\ustft) uses a sequential, data-adaptive mixture over a causal bank of candidate window durations. By scoring candidate windows with normalized order-three R\'enyi entropy and propagating state weights across a ternary (\emph{maintain}, \emph{grow}, \emph{reset}) Markovian trellis, the \ustft{} tracks evolving signal timescales in $\mathcal{O}(s_{\max})$ operations per frame while guaranteeing a universal mix-loss regret bound of $\mathcal{O}((k+1)\log M)$ relative to any hindsight schedule with $k$ resolution changes. Linearity of the Fourier transform ensures that the posterior mixture of candidate spectra corresponds to a single STFT computed with a data-adaptive effective window, and envelope-normalized overlap-add synthesis preserves exact time-domain reconstruction. The \ustft{} approaches the performance of hindsight optimal window length schedules.

\appendices

\section{Proof of Theorem~\ref{thm:entropy-floor}}
\label{app:stft-proof}

Fix a candidate support $L_s \in \mathcal{L}$ with $L_s \ge 2$, and let $\mathbb{K} = \C$ for the full complex DFT or $\mathbb{K} = \R$ for the unique-bin real DFT. Let $\iota_{L_s} : \mathbb{K}^{L_s} \to \mathbb{K}^N$ denote the linear operator that zero-pads a length-$L_s$ vector on the left to align with the rightmost $L_s$ coordinates of $\mathbb{K}^N$, and let $\mathcal{P}_{\mathcal{K}}$ project onto the retained frequency bins $\mathcal{K}$. Define the composite analysis operator
\begin{equation}
\mathcal A_{L_s}
=\mathcal P_{\mathcal K}\mathcal F_N\iota_{L_s}.
\end{equation}
Because $\mathcal{F}_N$ is invertible, $\iota_{L_s}$ has full column rank, and $\mathcal{P}_{\mathcal{K}}$ retains all non-redundant DFT bins under conjugate symmetry for $\mathbb{K} = \R$ (or all $N$ bins for $\mathbb{K} = \C$), $\mathcal{A}_{L_s}$ is injective: $\mathcal{A}_{L_s} z \ne 0$ for every $z \in \mathbb{K}^{L_s} \setminus \{0\}$.

By scale invariance of $p_{m,s}$ and $q_{m,s}$, it suffices to restrict $z$ to the compact unit sphere
\begin{equation}
\mathbb S_{L_s}
=\{z\in\mathbb K^{L_s}:\|z\|_2=1\}.
\end{equation}
For $z \in \mathbb{S}_{L_s}$, define the temporal and spectral PMFs
\begin{align}
p_j(z)&=|z[j]|^2,
\qquad 0\le j<L_s,\\
q_k(z)&=
\frac{\alpha_k|[\mathcal A_{L_s}z][k]|^2}
{\sum_{u\in\mathcal K}\alpha_u|[\mathcal A_{L_s}z][u]|^2},
\quad k\in\mathcal K.
\end{align}
Because $\mathcal{A}_{L_s}$ is injective on $\mathbb{S}_{L_s}$ and $\alpha_k > 0$, the denominator of $q_k(z)$ is strictly positive and continuous on $\mathbb{S}_{L_s}$. Consequently, the normalized joint resolution objective
\begin{equation}
\Phi_{L_s}(z)
=\frac{\Hthree(p(z))}{\log_2 L_s}
+\frac{\Hthree(q(z))}{\log_2 F}
\label{eq:proof-objective}
\end{equation}
is continuous on the compact set $\mathbb{S}_{L_s}$ and attains its global minimum by the Extreme Value Theorem.

To prove that this minimum is strictly positive, note that for any PMF $v$, $\sum_i v_i^3 \le \sum_i v_i = 1$, with equality if and only if $v$ is a Kronecker delta (point mass):
\begin{equation}
\Hthree(v)=0
\iff v\ \text{is a point mass}.
\label{eq:zero-renyi-characterization}
\end{equation}
Suppose for contradiction that $\Phi_{L_s}(z) = 0$ for some $z \in \mathbb{S}_{L_s}$. Nonnegativity of $\Hthree$ implies $\Hthree(p(z)) = 0$ and $\Hthree(q(z)) = 0$, so both $p(z)$ and $q(z)$ must be point masses. For $p(z)$ to be a point mass, $z$ must be concentrated at a single sample index $j$:
\begin{equation}
z=e^{i\theta}e_j
\end{equation}
(with $e^{i\theta} \in \{\pm 1\}$ when $\mathbb{K} = \R$). However, the DFT of a single impulse has flat magnitude across all frequency bins:
\begin{equation}
|[\mathcal A_{L_s}e_j][k]|^2
=\frac1N,\qquad k\in\mathcal K,
\end{equation}
which yields the strictly positive spectral PMF
\begin{equation}
q_k(e_j)
=\frac{\alpha_k}{\sum_{u\in\mathcal K}\alpha_u}>0,
\qquad \forall k\in\mathcal K.
\end{equation}
Because $F = |\mathcal{K}| \ge 2$, $q(e_j)$ has at least two strictly positive coordinates and cannot be a point mass, contradicting \eqref{eq:zero-renyi-characterization}. Thus,
\begin{equation}
c_{L_s}
\triangleq\min_{z\in\mathbb S_{L_s}}\Phi_{L_s}(z)>0.
\end{equation}
Taking the minimum over the finite candidate bank $\mathcal{L}$ yields
\begin{equation}
c_{N,\mathcal L,\mathcal K,\boldsymbol\alpha}
=\min_{L_s\in\mathcal L}c_{L_s}>0,
\end{equation}
completing the proof.

\bibliographystyle{IEEEtran}
\bibliography{refs}
\end{document}